# A Hybrid model for the origin of photoluminescence from Ge nanocrystals in SiO$_2$ matrix

A. Singha and A. Roy[*]

*Department of Physics and Meteorology,*

*Indian Institute of Technology Kharagpur, West Bengal 721302, India*

D. Kabiraj and D. Kanjilal

*Inter-University Accelerator Centre,*

*Aruna Asaf Ali Marg, New Delhi 110067, India*

## Abstract

In spite of several articles, the origin of visible luminescence from germanium nanocrystals in SiO$_2$ matrix is controversial even today. Some authors attribute the luminescence to quantum confinement of charge carriers in these nanocrystals. On the other hand, surface or defect states formed during the growth process, have also been proposed as the source of luminescence in this system. We have addressed this long standing query by simultaneous photoluminescence and Raman measurements on germanium nanocrystals embedded in SiO$_2$ matrix, grown by two different techniques: (i) low energy ion-implantation and (ii) atom beam sputtering. Along with our own experimental observations, we have summarized relevant information available in the literature and proposed a *Hybrid Model* to explain the visible photoluminescence from nanocrystalline germanium in SiO$_2$ matrix.

[*]Electronic address: anushree@phy.iitkgp.ernet.in

## I. INTRODUCTION

Though widely in use for optoelectronic device fabrication, Group IV in-direct band gap semiconductors, silicon (Si) and germanium (Ge), are very inefficient light emitters even at liquid helium temperature. However, it is possible to improve the efficiency of the light emission from these materials by changing their structure. For example, in superlattice structures, the size of the Brillouin zone is reduced and the bottom of the conduction band is folded onto the $\Gamma$ point. This changes the system to a direct gap material. For nanostructures, like individual quantum dots, the electronic states are completely discrete. The optical matrix element between pair of these states determines the probability of electronic transitions. For large nanocrystals, the Fourier component of the envelope functions of the confined carriers at the wavevector corresponding to the indirect gap transition is appreciable. These causes play the same role as the phonon does for the indirect transition in corresponding bulk materials. In the above mentioned types of modified structures (superlattices, nanostructures, large nanoparticles), the breakdown of momentum selection rules enhances the interband optical matrix element and hence, increases the luminescence efficiency. The low rate of non-radiative recombination also plays an important role in enhancing their light output [1]. In addition, the quantum confinement of charge carriers in semiconductor nanostructures increases their electronic band gap and hence results in a visible luminescence in these materials. There is a blue shift in the luminescence peak energy with decrease in size of the particles.

In recent years, the visible photoluminescence (PL) from Ge clusters, of few tens of nanometers or even lesser in size, grown in $SiO_2$ matrix, have attracted great interest due to their potential applications in improving the field of quantum optoelectronic devices.

However, our understanding of the origin of PL in this system is incomplete till date [2]. Some authors attribute the modified electronic structure of Ge-clusters due to quantum confinement to be the origin of visible luminescence from this system. The alternative model proposes that the visible luminescence is extrinsic to Ge clusters and originates from the defect states formed during the growth process. Below we review, in brief, the present understanding on the above issue.

**Photoluminescence in Ge nanocrystallites**

In the last two decades, the visible PL from nanocrystalline Si and Ge, has been studied extensively. We start with a comparison of a few bulk properties of these two elements in the periodic table. The static relative dielectric constant of bulk Ge is 15.8 and that of bulk Si is 11.7 [3]. In addition, the light and heavy holes in bulk Ge have masses $\sim 0.043m$ and $\sim 0.34m$ and in silicon $\sim 0.16m$ and $\sim 0.52m$, where m is the mass of the free electron [4]. Because of the higher dielectric constant and smaller carrier mass in Ge than in Si, one can expect the effect of quantum confinement of charge carriers in the electronic structure to be more pronounced in Ge than in Si nanostructures [5, 6]. Moreover, as the direct band gap 0.88 eV of Ge is close to the indirect band gap 0.75 eV, the Ge-nanostructures are expected to exhibit a direct-gap semiconductor nature [5, 6]. It is also to be noted that the band gap of Ge nanocrystals is sensitive to the size of the crystals [5].

Visible luminescence at around 2.2 eV from nanocrystalline Ge (NC-Ge) in $SiO_2$ matrix, grown by rf co-sputtering deposition technique, was first reported by Maeda et al [7, 8, 9] and was explained by quantum confinement of the charge carriers in the system by simultaneous, high resolution transmission electron microscopic, X-ray photoelectron (XP), Raman, and PL spectroscopic measurements. The blue shift in emission energy from 0.88 eV to 2.0 eV with decrease in size of the particles in a similar system, indicating the recombination of confined electron-hole pairs as the origin of PL in NC-Ge, has been reported by Takeoka et al in [10]. However, there are many experiments reported later, where it has been shown through different measurements that the quantum confinement cannot be taken as the origin of visible PL in NC-Ge in $SiO_2$ matrix. Here, we mention some of these results. Min et al have shown that the strong luminescence peak energy at around 1.8 eV and the decay lifetime from NC-Ge in $SiO_2$ matrix, prepared by ion implantation technique and then annealed in high vacuum at different annealing temperatures (600 °C - 1000 °C), exhibit a poor correlation with the size of the nanocrystals when compared to radiative recombination of quantum confined excitons in Ge quantum dots. They attributed the visible PL to the defect related states in $SiO_2$ matrix and confirmed their conjecture by showing that the $Xe^+$ implanted $SiO_2$ sample also exhibits the similar luminescence behavior [11]. Samples prepared by similar technique (annealed in nitrogen atmosphere) have also been studied by Kim et al using simultaneous XP spectroscopic measurements and PL measurements [12]. The blue luminescence, at around 3 eV, from as-implanted NC-Ge in $SiO_2$ matrix has been correlated with defects at the nanocrystal/matrix interface or in the matrix itself, formed during implantation. However, from the significant increase in intensity of Ge-Ge bonds, as observed from XP spectroscopic measurements, the authors have attributed the origin of the PL peak from the annealed samples to the quantum confinement effect in NC-Ge. From infra-red spectroscopic measurements and electron spin resonance it has been suggested by Wu et al that PL bands (at 1.7 and 1.9 eV) in NC-Ge in porous silicon structure, prepared by pulsed laser deposition of Ge on porous Si, originates from optical transitions in the oxygen related defect centers at the interface between porous silicon and NC-Ge [13]. The variation in electron spin density of Si-O/Si-H bond has been directly correlated with the PL intensity from the sample. The ab-initio electronic structure calculation by Niquet et al [14] for NC-Ge using sp$^3$ tight binding description presented an analytical law : (a) the blue-green luminescence (above 2 eV) of NC-Ge crystallites originates from the defects in the oxides and (b) the observed size dependent PL energy (as mentioned in [7, 8, 9, 10]) near the infra-red region may be due to the deep trap in the band gap of the nanocrystallites. The above confusion was supplemented with the fact that unlike nanocrystalline Si for NC-Ge the oscillator strength of electronic transitions are quite high at the absorption edge, even for dots with a diameter of about 2.2 nm [15]. Thus, it is difficult to rule out the possibility of getting strong luminescence from NC-Ge itself (as could be done in explaining the origin of visible PL from Si nanostructures [15]).

From the above brief discussion we see that the understanding of the origin of PL from NC-Ge in SiO$_2$ matrix is still not very clear. In this article, we have addressed this long standing query by carrying out simultaneous PL and Raman measurements on NC-Ge in SiO$_2$ matrix grown by two different routes (i) by ion implantation and (ii) by atom beam sputtering. The motivations behind choosing these two different techniques are the following : In semiconductor-based devices, ion-implantation is an attractive technique for selective-area doping, precise control over dopant concentration etc. The main disadvantage of this technique is that the ion irradiation knocks out atoms from the irradiated material and creates point defects in the lattice. These defect states can interact with light and participate in absorption or emission of photons in radiative/nonradiative recombination processes. On the other hand, in the atom beam sputtering technique, the 1.5 keV neutral Ar atom beam is directed towards the target to be sputtered in the vacuum chamber [16]. A saddle field source, with DC power, provides a medium area beam of essentially neutral atoms. A great variety of materials, including alloys, compounds, pure metals and mixtures, can be sput-tered. The electronics industry are the frequent users of sputter deposition because of the advantages of thickness control, high density, composition reproducibility, and automation. The main disadvantage of this technique includes its lower deposition rates. In this technique one does not expect to have defects, like vacancies (as in case of ion implantation), in the grown sample. However, the presence of oxide related defect states in the samples cannot be ruled out. It is also to be noted that the growth techniques, ion implantation and co-sputtering, have been widely used in the literature for preparation of NC-Ge in $SiO_2$ matrix.

We have probed the origin of PL in the samples by simultaneous PL and Raman spectroscopic measurements, as Raman spectroscopy is a non-destructive probe to identify chemical structures in a sample. In the present case, this technique is capable of providing the fingerprint evidence for the formation of Ge-Ge bonds and the qualitative information about the other chemical structures in the sample with a particular optical property. Raman line shape can also be used to determine the size of the nanocrystals.

The Section II of this article covers the sample preparation techniques following the above two routes and other details regarding the instruments, which we have used for PL and Raman measurements. In Section III, we have presented our experimental results obtained from Raman and PL spectroscopic measurements, in detail. Section IV discusses our own results and observations by other authors, available in the literature, and proposes a new model to explain visible luminescence in NC-Ge in $SiO_2$ matrix.

## II. EXPERIMENTS

### A. Sample preparation

*1. Preparation of sample by ion-implantation:*

A SiO$_2$ film of thickness 100 nm was thermally grown by wet oxidation on a p-type Si(100) wafer and used as a matrix. $^{74}$Ge$^+$ ions with the energy 150 keV were implanted in this matrix at room temperature. The employed dose of the ion was 3×10$^{16}$. In this article, the as-implanted sample is referred to as Sample 1A. Subsequently, the sample was annealed in N$_2$ atmosphere for 30 mins. at different temperatures. Five samples, Sample 1B- Sample 1F, were grown by varying the annealing temperatures (500, 600, 800, 950, and 1100° C).

*2. Preparation of sample by atom beam sputtering*

A composite of Ge and SiO$_2$ was sputtered by 1.5 kV Argon atom beam onto Si by atom beam co-sputtering and then annealed in forming gas (96% Ar+ 4% H$_2$) for 30 mins. at temperatures 400, 600 and 800°C. In this article, we will refer to the as-grown sample as Sample A and the samples annealed at the above temperatures as Samples 2B, Sample 2C, and Sample 2D, respectively.

### B. Measurement Techniques

Raman spectra of all our samples were measured in back-scattering geometry using a 488 nm Argon ion laser as an excitation source. The spectrometer was equipped with a 1200 grooves/mm holographic grating, a holographic super-notch filter, and a Peltier cooled CCD detector. With 100 $\mu$m slit-width of the spectrometer the resolution of our Raman measurement was 1.0 cm$^{-1}$. Photoluminescence spectra were also recorded using the same

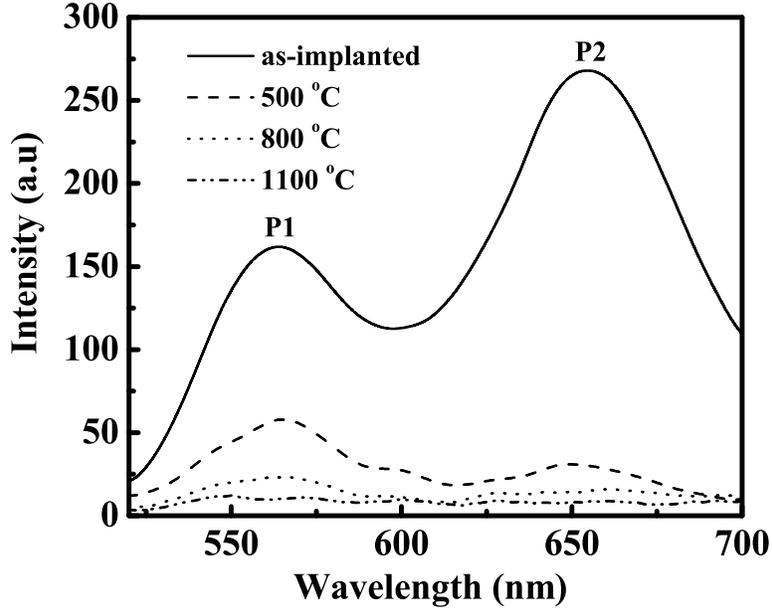

FIG. 1: PL spectra of as-implanted sample, Sample 1A, and samples annealed at 500 °C (Sample 1B), 800 °C (Sample 1D) and 1100 °C (Sample 1F).

spectrometer (without notch filter) and the same excitation source.

### III. RESULTS

#### A. Raman and photoluminescence measurements for Sample 1A - Sample 1E

From TRIM calculation [17] we find that during Ge ion implantation with energy 150 keV and dose $3 \times 10^{16}$ into 100 nm $SiO_2$ matrix, the recoil atoms lead to a dissociation of the oxide to O and Si. In general, these atoms react very fast and form $SiO_2$. However, a small fraction of atoms remain unbound and diffuse in the matrix. The unbound O atoms diffuse faster towards $Si/SiO_2$ interface. Thus, above this interface region the concentration of unbound Si increases. In other words, at the end of the implantation process, there is an excess of Si above the interface. During annealing these Si atoms agglomerate and act as nucleation centers for diffusing Ge from the implanted $SiO_2$ region [14].

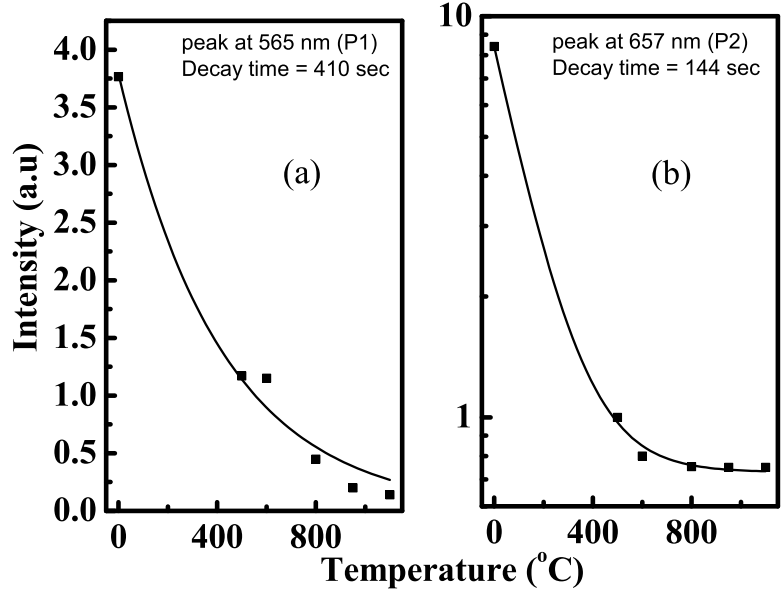

FIG. 2: Change in intensities of (a) peak P1 at 565 nm and (b) peak P2 at 657 nm with annealing, from Sample 1A- Sample 1F. Note the log scale in (b).

Fig. 1 shows the room temperature PL spectrum of as-implanted sample (Sample 1A) and a few characteristic PL spectra of post-annealed implanted samples (Sample 1B, Sample 1D, and Sample 1F). Sample 1A exhibits two strong peaks, one at 565 nm (2.2 eV), P1, and the other at 657 nm (1.9 eV), P2. The luminescence intensity of peak P1 decreases steadily with annealing [Fig. 2(a)]. The intensity of P2 decreases 10 times when the sample is annealed at 500 °C and decreases more with further annealing [Fig. 2(b)].

Fig. 3 shows the Raman spectra of Sample 1A to Sample 1F. We try to deconvolute each spectrum with different components, to obtain the information of the change in chemical characteristics of the samples with annealing. In all spectra, we observe two prominent peaks, one appears at $\sim 300$ $cm^{-1}$ and the other at $\sim 430$ $cm^{-1}$. A broad feature at 350 $cm^{-1}$ is also present in all spectra as a high frequency shoulder of the peak at 300

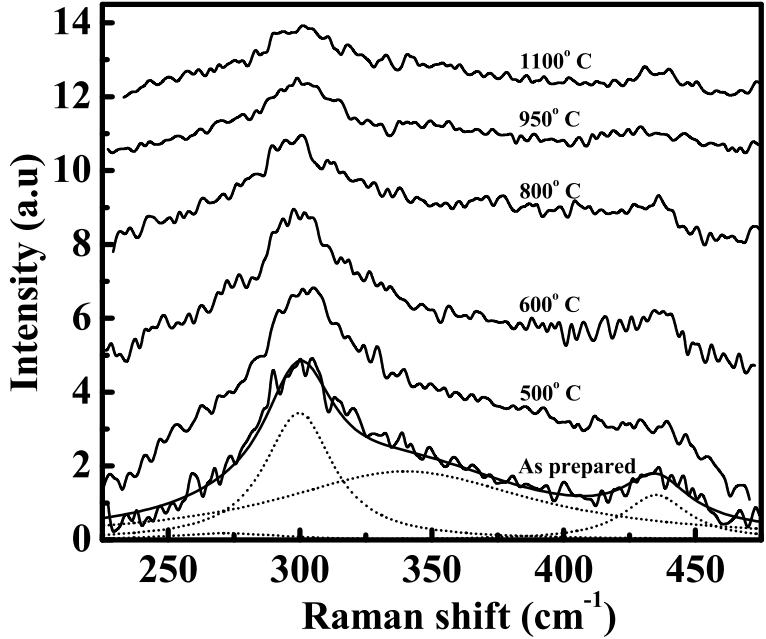

FIG. 3: Raman spectrum of as-implanted sample (Sample 1A) and samples annealed at different annealing temperatures (Sample 1B - Sample 1F.)

cm$^{-1}$. The allowed first order $\Gamma$ point phonon mode of bulk Ge is expected to appear at 302.6 cm$^{-1}$ [18]. In NC-Ge, it is expected to be slightly shifted and asymmetric towards the low frequency side due to confinement of phonon, discussed later. It is to be noted that the two-phonon density of states of bulk Ge shows a feature at around 350 cm$^{-1}$ [19]. Being a higher order scattering process, this particular feature is nearly absent in case of bulk Ge. However, in a nanocrystalline environment, due to breakdown of selection rules, the forbidden/weak vibrational modes can contribute appreciably to the first order Raman scattering. We attribute the observed feature at 350 cm$^{-1}$ to the second order Raman mode from NC-Ge. The feature at around 430 cm$^{-1}$ can be due to the presence of Si-Ge bond or Si$_x$Ge$_{1-x}$ alloy in implanted layers. In Ref. [20] it was argued that both features at 300 cm$^{-1}$ and 430 cm$^{-1}$ in such samples can be two-photon Raman peaks from the Si substrate on which the SiO$_2$ layer is grown. In our samples, the intensities of these peaks change with annealing, which is not expected if they arise from Si substrate (the intensity of the Si peak at 522 cm$^{-1}$ remains unchanged with annealing). Keeping in mind, the possibility of the presence of amorphous Ge (a-Ge)/GeO$_x$ in the samples, we have also included a feature at 274 cm$^{-1}$ in our analysis. We made an attempt to analyze each spectrum by deconvoluting it with the above-mentioned four features : at $\sim$ 274 cm$^{-1}$ due to a-Ge/GeO$_x$, at $\sim$300 and 350 cm$^{-1}$ due to NC-Ge and at 430 cm$^{-1}$ due to Si-Ge bond or Si$_x$Ge$_{1-x}$ in the sample. These four components are shown by dashed lines in Fig. 3 only for the as-implanted sample.

In bulk crystals, the phonon eigenstate is a plane wave and the wavevector selection rule requires q$\approx$0. In contrast, for nanoparticles the spatial correlation function of the phonon becomes finite due to its confinement in the nanocrystal and hence the q$\approx$0 selection rule is relaxed [21]. The Raman spectrum $I(\omega)$ due to this confined optic phonon is given by,

$$I_c(\omega) = \int \frac{d\mathbf{q}|C(0,\mathbf{q})|^2}{[\omega - \omega(\mathbf{q})]^2 + (\Gamma_0/2)^2}, \tag{1}$$

where, $\omega(\mathbf{q})$ and $\Gamma_0$ are the phonon dispersion curve and the natural line width (FWHM) of the corresponding bulk materials, $C(0,\mathbf{q})$ is the Fourier coefficient of phonon confinement function. For nanoparticles, it has been shown that the phonon confinement function, which fits the experimental data best is $W(\mathbf{r}, L) = \exp\left(\frac{-8\pi^2 r^2}{L^2}\right)$, the square of the Fourier coefficient of which is given by $|C(0,\mathbf{q})|^2 \cong \exp -\frac{q^2 L^2}{16\pi^2}$. $L$ is the size of the particles. The integration in Eqn.1 must be performed over the whole Brillouin zone. The average phonon dispersion in the Ge crystal for the LO phonon modes is taken as [22]

$$\omega(q) = \omega_0 - (\Delta\omega)^2, \tag{2}$$

which fits the experimental curve well in the direction of Γ-M upto $q_{max} = 0.456$. $\omega_0$=302.6

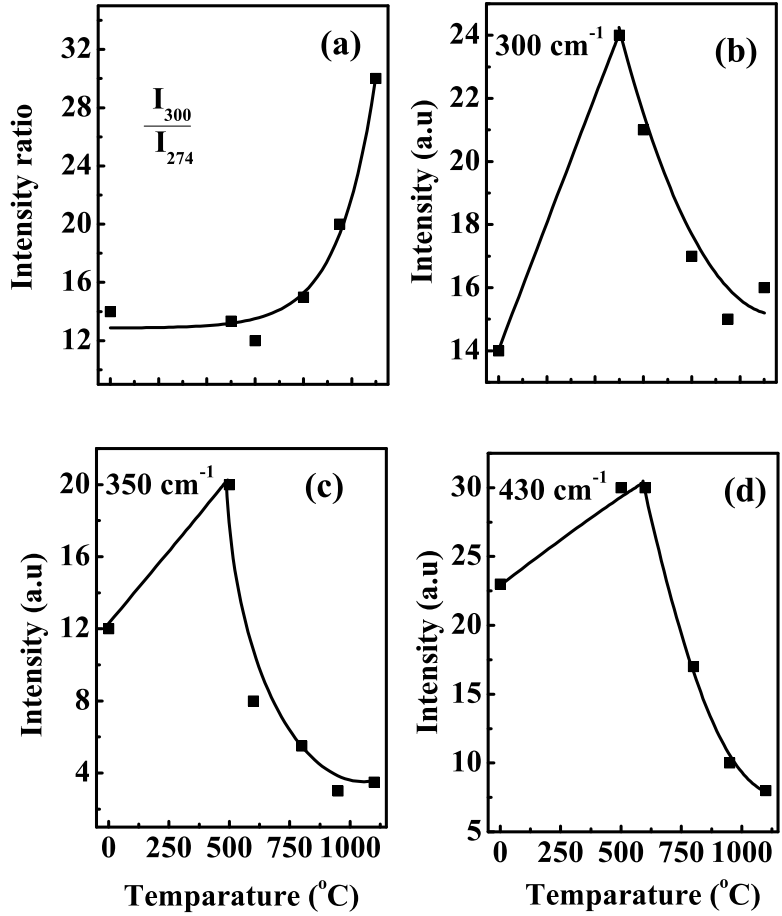

FIG. 4: Change in (a) the ratio of the intensities of the Raman peaks at 300 cm$^{-1}$ and 274 cm$^{-1}$ and the intensity of the Raman peak at (b) 300 cm$^{-1}$, (c) 350 cm$^{-1}$ and (d) 430 cm$^{-1}$ with annealing from Sample 1A- Sample 1F.

cm$^{-1}$ is the corresponding bulk LO phonon frequency. $\Delta\omega$ is the band width of bulk LO phonon branch : $\Delta\omega$ =57 cm. In the above equations, $q$ is taken in units of $2\pi/a$, where, $a$ is the lattice constant of the material. We have taken $a$=5.66 Å[23].

Following the above discussion, we fit each spectrum with a combined Raman line shape $I = I_c + I_{274} + I_{350} + I_{430}$. $I_{274}$, $I_{350}$ and $I_{430}$ are Lorentzian line shapes for the features at 274, 350, and 430 cm$^{-1}$. We have kept the intensities, peak positions and widths as free fitting parameters. The variations in intensities of these four peaks with annealing from Sample 1A to Sample 1F are shown in Fig. 4(a) - Fig. 4(d). From this fitting procedure, the particle size, $L$, in as-implanted sample is estimated to be 5±2 nm, which remains nearly unchanged for all samples. The ratio of the intensity of the peak due to NC-Ge at ∼ 300 cm$^{-1}$ and a-Ge at 274 cm$^{-1}$, $I_{300}/I_{274}$, increases with annealing indicating a relative increase in crystalline Ge in annealed samples due to precipitation and simultaneous decrease in amorphous GeO$_2$ in the material, see Fig. 4(a). From the Gibbs energy at the thermodynamical equilibrium of the reaction, it can be shown that Ge is very stable in the SiO$_2$ network [7]. The thermodynamic reaction between GeO$_2$, Si, SiO$_2$ and Ge is expected to be [7]

$$GeO_2 + Si = SiO_2 + Ge$$

The above reaction can proceed spontaneously till 300 °C. However, it is to be noted that GeO$_x$ is volatile particularly at temperatures higher than 400 °C [24]. If we study the intensity variation of the individual features due to a-Ge and NC-Ge at 274 and 300 cm$^{-1}$, there is an initial increase of their intensities, followed by a decrease with further annealing [Fig. 4(b) for NC-Ge]. The intensity of the peak at 340 and 430 cm$^{-1}$ show nearly the same trend [Fig. 4(c) and 4(d)]. The decrease in intensity of the overall spectrum with annealing beyond 500 °C is due to decrease in film thickness as a result of evaporation of GeO$_x$ during high temperature annealing.

To understand the origin of the two PL peaks, P1 and P2 in implanted samples, we now try to correlate the variation in intensities of these peaks, shown in Fig. 2(a) and Fig. 2(b), with change in intensity of Raman bands, shown in Fig. 4(a) to Fig. 4(d) with annealing temperature. None of the features in the PL spectra shows the same trend with annealing as observed for intensities of different Raman lines in the samples. Also, it is to be noted that annealing the samples in nitrogen atmosphere (i) does not passivate defect states and (ii) results in thermal relaxation of the matrix. Thus, the monotonous decrease in intensity of the peak P1 in PL spectra, in Fig. 2(a), can be due to thermal relaxation of the defect-related states with annealing. The intensity of P2 in the sample annealed at 500 ° is ten times less than the same in as-implanted sample and then decays exponentially with further annealing. It is difficult to attribute such a sudden drop in intensity in annealed sample to the relaxation of the matrix by nitrogen. If one looks carefully, the exponential decay rate of P2 (144 sec) is very close to exponential growth rate (152 sec) of $I_{300}/I_{274}$ in our sample. Thus, though the PL peak at 565 nm is originating from the defect related states, it is more likely that the PL peak P2 at 657 cm$^{-1}$ is related to both NC-Ge and a-Ge components in the implanted samples.

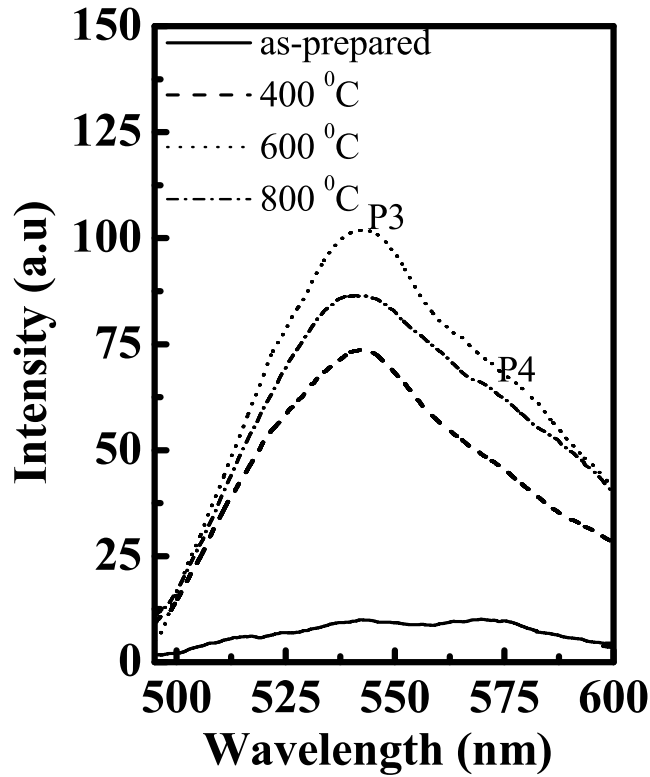

FIG. 5: PL spectra of Sample 2A - Sample 2D.

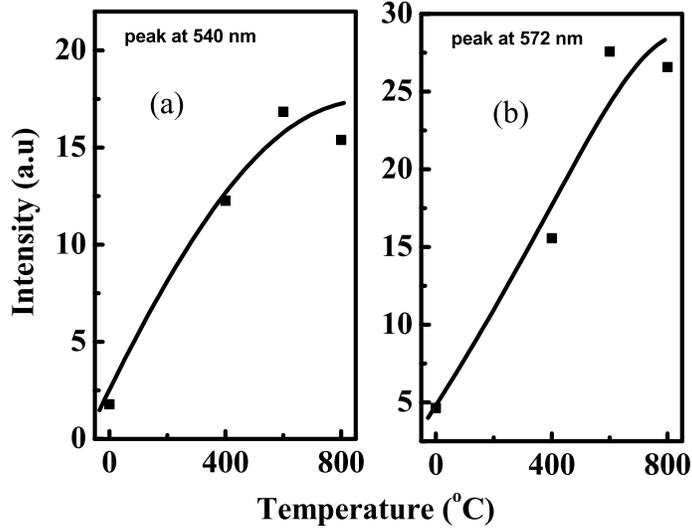

FIG. 6: Change in intensities of (a) peak P1 at 540 nm and (b) peak P2 at 572 nm with annealing, from Sample 2A to Sample 2D.

## B. Raman and photoluminescence measurements for Sample 2A - Sample 2E

Fig. 5 shows the room temperature PL spectra of the co-sputtered samples, pristine sample (Sample 2A) and samples annealed at different temperatures, Sample 2B- Sample 2D. Each sample exhibits a broad PL spectrum with two luminescence features. The peak (P3) appears at 540 nm (2.3 eV). The other feature (P4), at 572 nm (2.17 eV), appears as a shoulder of P3. The intensity of both features increase with annealing [Fig. 6].

Fig. 7 shows the Raman spectra of Sample 2A to Sample 2D. The Raman line shape for these samples carry all basic features, as observed in the same for the implanted sample [in Fig. 3]. The Raman spectra of these co-sputtered samples are narrower. The reasons can be (i) less number of non-radiative centers and/or (ii) less size distribution of NC-Ge in co-sputtered samples than in implanted samples. Based on the arguments presented in Section III A, we deconvoluted each spectrum into four components. The variations in intensity of different components are shown in Fig. 8(a)- 8(c). Though, in Fig. 8(a), for these co-sputtered samples, we observe an increase in the intensity ratio of the peak at 300 and 274 cm$^{-1}$ with annealing, the nature of change is different compared to what we have observed in Fig. 4(a) for implanted samples. We propose that this change is due to different annealing condition for our implanted samples and co-sputtered samples. As shown in Fig. 8(b) and Fig. 8(c), the intensity of the peak at 300 and 350 cm$^{-1}$ due to NC-Ge decreases with annealing because of thinning of the film due to evaporation of GeO$_x$, as observed for implanted samples. The intensity of the peak at around 430 cm$^{-1}$ due to Si-Ge bond or Si$_x$Ge$_{1-x}$ is quite small. It is difficult to show its variation in intensity within experimental accuracy.

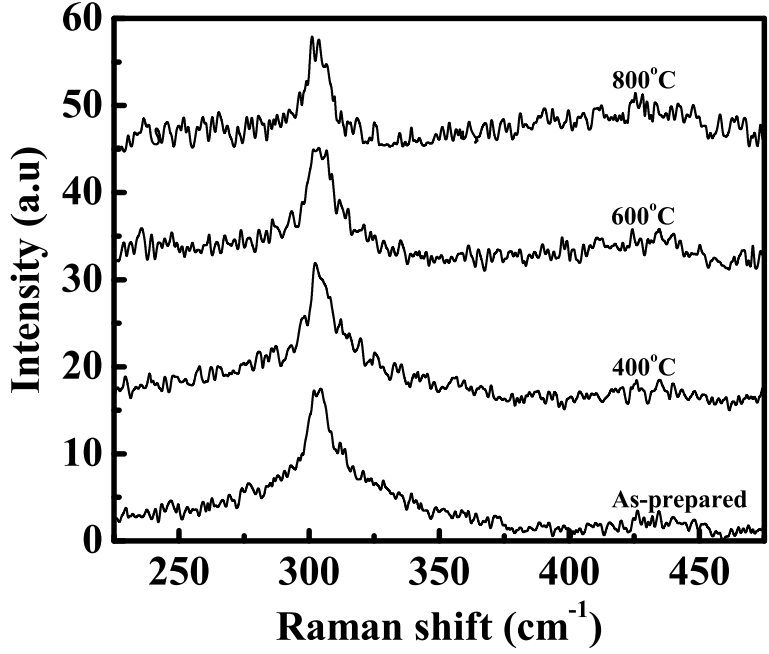

FIG. 7: Raman spectrum of Sample 2A - Sample 2D.

We now try to understand the origin of the PL peaks in co-sputtered samples. None of the features, either P3 or P4 in the PL spectra in Fig. 5, originates from defect states, as intensity of both peaks increases with $H_2$ annealing. While annealing under $N_2$ atmosphere causes thermal relaxation of the matrix, the annealing in $H_2$ helps to passivate non-radiative defect states. The increase in intensity of these features is mostly due to passivation of non-radiative defect states while annealing in $H_2$ atmosphere. This clearly indicates that NC-Ge or a-Ge or both of these components may be the origin of these PL peaks in co-sputtered samples.

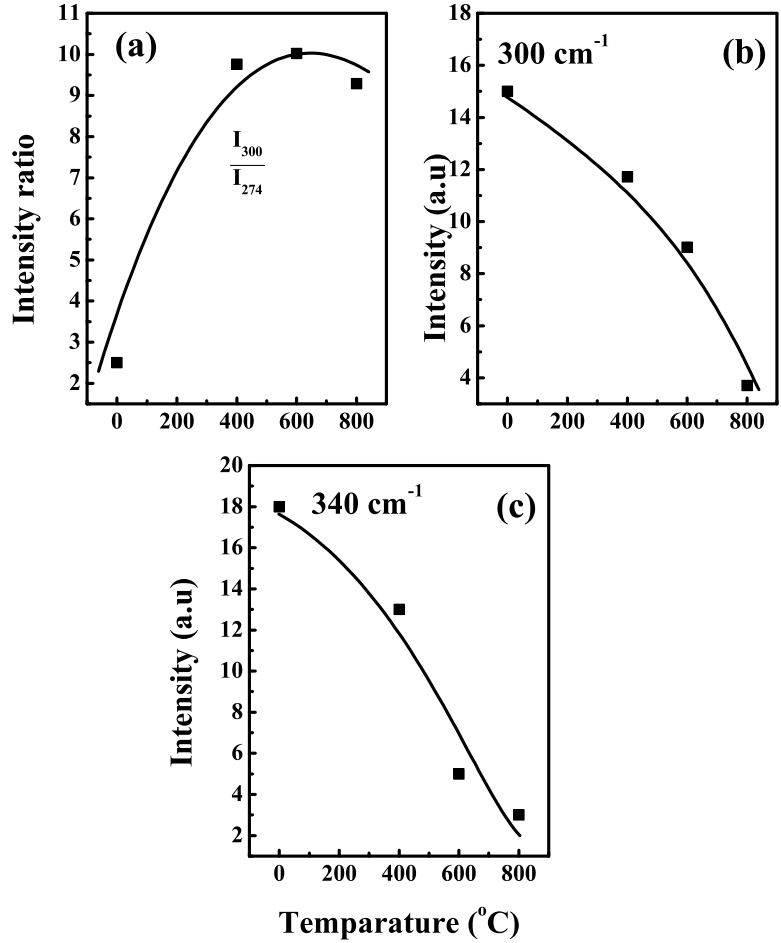

FIG. 8: Change in (a) the ratio of the intensities of the Raman peaks at 300 cm$^{-1}$ and 274 cm$^{-1}$ and the intensity of the Raman peak at (b) 300 cm$^{-1}$, (c) 350 cm$^{-1}$ with annealing from Sample 2A- Sample 2D.

## IV. DISCUSSION

From our observations, presented in the previous section, and the available literature we propose the following: the PL in near infra-red region from NC-Ge in $SiO_2$ matrix can have two possible origins (a) it can be due to defect states in the matrix and/or (b) it can be related to the NC structure and the oxide related defect/amorphous states, together, in the system. For the latter case our conjecture is: electron-hole pairs are generated within the nanocrystalline particles. However, they do not recombine within this structure, but diffuse out. The radiative recombination takes place at the oxide related amorphous states outside the crystalline structure. This model is similar to the hybrid model used to explain the visible PL in porous Si [25, 26].

We now discuss different arguments to support our model:

(1) We have observed strong visible PL and the presence of NC-Ge in all our samples. In implanted sample, we observed two PL peaks. The PL peak P1 in ion-implanted sample must have its origin in defect states only, as its intensity quenches steadily with annealing. Moreover, there is no correlation between the intensity variation of this particular PL band and any chemical component present in the sample, as observed from Raman measurements. On the other hand, as we have shown, the peak P2 in the implanted sample or peaks P3 and P4 in the co-sputtered sample, are not directly related to defect states. Rather it is more likely that they originate from NC-Ge and/or a-Ge related states in the samples. In other words, it is not always true that the PL peaks near IR/visible region originates from defect related states in the material as has been concluded in [11].

(2) We find a similar report by Okamoto and Kanematsu in [27], where, NC-Ge in $SiO_2$ matrix was prepared by rf cosputtering of Ge and $SiO_2$. From the difference in PL excitation and absorbtion spectra these authors have suggested that the photo-generated excitons in NC-Ge do not recombine within the same structure but do so at different sites. This has been further confirmed by the study of PL dynamics. The PL radiative decay rate has been estimated to be of order of nanosecond and independent of size of the crystallites. This observation is in contradiction to the expected PL decay behavior as has been shown by Takagahara and Takeda from electronic structure calculations using ab-initio methods [5]. According to the calculation, the radiative life time of excitons in NC-Ge is strongly size dependant. The fine structure of the PL spectrum, shown in Ref. [27], taken at 2K also cannot be explained by exciton-phonon coupling in solid Ge. It is more probable that the Frohlich interaction between excitons and the stretch vibrations of the surface species is the cause of the fine structures in the PL spectrum.

(3) Visible PL from NC-Ge formed by $H_2$ reduction of $Si_{0.6}Ge_{0.4}O_2$ has been reported by Paine et al in [28]. Here the authors have shown by Raman measurements that for annealing with longer time, there is decrease in oxide content in $H_2$ atmosphere, following the reaction $GeO_2 + H_{2(v)} = Ge + H_2O_{(v)}$, which systematically increases the Ge component in the sample. However, the PL spectra shows a non-monotonous behavior. Its intensity increases in the beginning, and then decreases with further annealing. However in [? ], the authors have left their observations unexplained. We believe that this result can be explained by our hybrid model. In as-grown sample, the absence of PL is due to the absence of NC-Ge in the system. During initial annealing in $H_2$ atmosphere, an optimum condition, when the presence of NC-Ge and GeO is sufficient for the sample to exhibit visible orange PL. However, with further annealing, there is a decrease in oxide content in the sample. At this stage, though the excitons are generated within the nanostructure, the radiative recombination channels through the outer oxide layer are nearly absent. This decreases the PL intensity.

(4) In Ref. [8], Maeda et al have studied visible PL at 2.18 eV from NC-Ge embedded in SiO$_2$ matrix. The samples were grown by rf sputtering technique and then annealed at 800 °C. They have shown the following: (a) a decrease in GeO$_2$ with an increase in NC-Ge component in the samples with annealing from XP spectroscopic measurements and (b) the luminescence only from the annealed sample. From these observations, they have concluded that the origin of luminescence is the excitonic confinement in NC-Ge. We note that these observations can also be explained by the hybrid model, proposed by us. The absence of luminescence in as-prepared sample in ref. [8] may be due to the absence of NC-Ge : as, according to our model, the excitons in this system, are generated within the nanostructures though they recombine radiatively outside . If one looks carefully, the Raman and XP spectra presented in this article by Maeda et al, one can clearly see the presence of a-Ge in the annealed sample. The need of both the presence of NC-Ge and a-Ge cannot be totally ruled out from the data presented in this article.

(5) A recent report on the structure and PL properties of GeO$_x$ on Si substrate shows that the PL peak of GeO$_x$ appears at $\sim$ 550 nm, which disappears when the sample is annealed at 400 °C in high vacuum [29]. However, with further annealing, a new PL peak appears at 1250 nm along with precipitation of NC-Ge. The PL radiative decay rate of 550 nm band has been estimated to be 20 ns, while that of the latter is of 200 ns. This PL behavior of GeO$_x$ does not match with ours/others observations on PL characteristics of NC-Ge in SiO$_2$ matrix. Thus we propose that the origin of visible PL in the system under our investigation, cannot be only due to GeO$_x$ in the samples.

(6) One may note that in most of the reports [11], where the luminescence from such a

system has been attributed to a-Ge/GeO$_2$, authors have also shown the presence of NC-Ge in the system. This, in an indirect way, suggests that the presence of both NC-Ge and amorphous states may be necessary for the NC-Ge in SiO$_2$ matrix to exhibit visible PL. However, we would like to point out that there is a possibility of strong luminescence only from the defect states, as has been seen observed by Min et al or in case of the PL peak P1 in our implanted sample. It is most likely that this kind of luminescence can only be observed from ion-implanted samples.

(7) A more controlled growth of Ge 'hut' clusters on Si(001) by molecular beam epitaxy technique, and Ge island on Si(001) by ultra-high vacuum chemical vapor deposition technique resulted in an emission spectrum at 0.85- 0.89 eV [30, 31]. Similar PL emission spectrum has been observed in the NC-Ge deposited by pulsed laser deposition in a Si matrix [32]. For these samples the luminescence from defect states can possibly be completely ruled out. Note that the PL energy in NC-Ge is much lower than the energy reported for NC-Ge in SiO$_2$ matrix.

In conclusion, we propose that for NC-Ge in SiO$_2$ matrix, the nanostructure serves as a source of electron-hole pairs, which recombine at radiative defect related states outside the nanostructure. A hybrid model only, thus can possibly explain the visible PL in this system.

## V. ACKNOWLEDGEMENTS

A. Roy thanks Department of Science and Technology, India for financial support.